\documentclass[aps,prl,reprint,superscriptaddress,floatfix,amsmath,showpacs]{revtex4-1}

\usepackage{graphicx}
\usepackage{dcolumn}
\usepackage{bm}
\usepackage{times}
\usepackage{textcomp}
\usepackage{amsmath}

\begin{document}
\preprint{Preprint}

\title{Stimulated Thermalization of a Parametrically Driven Magnon Gas \\ 
as a Prerequisite for Bose-Einstein Magnon Condensation}

\author{P. Clausen}
\email{Clausen@physik.uni-kl.de}
\affiliation{Fachbereich Physik and Landesforschungszentrum OPTIMAS, Technische Universit\"at Kaiserslautern, 67663 Kaiserslautern, Germany}

\author{D. A. Bozhko}
\affiliation{Fachbereich Physik and Landesforschungszentrum OPTIMAS, Technische Universit\"at Kaiserslautern, 67663 Kaiserslautern, Germany}
\affiliation{Graduate School Materials  Science in Mainz, Gottlieb-Daimler-Stra{\ss}e 47, 67663 Kaiserslautern, Germany}

\author{V. I. Vasyuchka}
\affiliation{Fachbereich Physik and Landesforschungszentrum OPTIMAS, Technische Universit\"at Kaiserslautern, 67663 Kaiserslautern, Germany}

\author{B. Hillebrands}
\affiliation{Fachbereich Physik and Landesforschungszentrum OPTIMAS, Technische Universit\"at Kaiserslautern, 67663 Kaiserslautern, Germany}

\author{G. A. Melkov}
\affiliation{Faculty of Radiophysics, Taras Shevchenko National University of Kyiv, 01601 Kyiv, Ukraine}

\author{A. A. Serga}
\affiliation{Fachbereich Physik and Landesforschungszentrum OPTIMAS, Technische Universit\"at Kaiserslautern, 67663 Kaiserslautern, Germany}

\date{\today}

\begin{abstract}

Thermalization of a parametrically driven magnon gas leading to the formation of a Bose-Einstein condensate at the bottom of a spin-wave spectrum was studied by time- and wavevector-resolved Brillouin light scattering spectroscopy. It has been found that the condensation is preceded by the conversion of initially pumped magnons into a second group of frequency degenerated magnons, which appear due to parametrically stimulated scattering of the initial magnons to a short-wavelength spectral region. In contrast to the first magnon group, which wavevectors are orthogonal to the wavevectors of the magnons at the lowest energy states, the secondary magnons can effectively scatter to the bottom of the spectrum and condense there.

\end{abstract}


\maketitle

Parallel parametric pumping \cite{Schloemann1960, Rezende1990}, where an alternating magnetic pumping field $\mathbf{h}(t)$ is applied parallel to the direction of a bias magnetic field $\mathbf{H}$, can effectively couple the spin system of a magnetic material with an external microwave energy source, and thus is widely used to generate \cite{Demokritov2003, Braecher2011}, amplify \cite{Kalinikos1994, Bagada1997, Braecher2014}, and restore \cite{Melkov2001, Chumak2014} spin-wave signals in macro- and micro-sized magnetic structures. In addition, this technique fruitfully serves for experimental studies of nonlinear dynamics in multi-mode wave systems as well. One of the greatest achievements in this area was the realization of a Bose-Einstein condensation of magnons \cite{Demokritov2006}, where the parametrically pumped magnon gas undergoes a phase transition leading to the spontaneous formation of a coherent state at the bottom of the spin-wave spectrum \cite{Pokrovsky2013}. Since the moment of this discovery, the physics of the parametrically driven magnon gases is under active theoretical and experimental investigation. In particular, the time dependent behavior of the magnon gas in the phase-energy space \cite{Demidov2007, Demidov2008-1, Demidov2008-2, Chumak2009}, the temperature of the magnon gas \cite{Serga2014}, and the role of different scattering mechanisms \cite{Hick2012} in the process of the thermalization of the injected magnons are in the focus of attention. 

\begin{figure}
  \includegraphics[width=0.95\columnwidth]{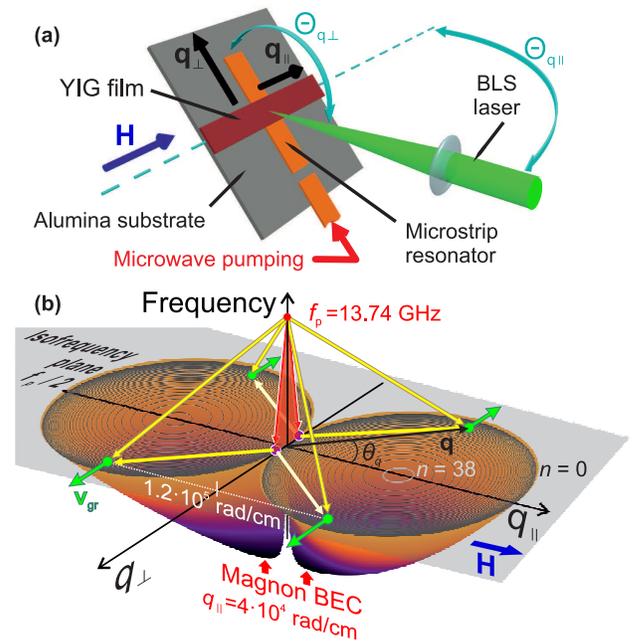}
  \caption{\label{Fig1}(color online). (a) Sketch of the experimental setup.
$\bf{H}$ is a bias magnetic field. $\bf{q}_{\parallel}$ and $\bf{q}_{\perp}$ are in-plane wavevectors of the transversal ($\bf{q}\perp\bf{H}$) and longitudinal ($\bf{q}\parallel\bf{H}$) magnons. The setup is tiltable by the angles $\Theta_{q\perp}$ and $\Theta_{q\parallel}$ allowing for wavevector resolution. (b) The magnon spectrum calculated for the first 38 thickness modes $n$ \cite{Kalinikos1986}. The shadowed area between the red arrows illustrates the magnon injection by parametric pumping. The initially pumped magnon group is marked by two violet dots. The second magnon group is shown by four green dots. The green arrows indicate group velocities of the secondary magnons. The yellow arrows outline the parametrically stimulated magnon scattering.
}
\end{figure}

Here, we provide experimental insight into the evolution of a magnon gas affected by four-magnon scattering in the presence of an external pumping field. We claim that the magnons initially pumped to the transversal branch of the spin-wave spectrum (magnon wavevector $\mathbf{q} \perp \mathbf{H}$) do not scatter into the lowest energy states located at the longitudinal branch ($\mathbf{q} \parallel \mathbf{H})$. Such a scattering and the consequent formation of a Bose-Einstein condensate (BEC) are associated with the parametric excitation of a second group of short-wavelength magnons propagating at an angle $\angle{(\mathbf{q},\mathbf{H})} < 90^\circ$. Simultaneously, the initially excited transversal magnon group is suppressed to the near thermal level by a mutual action of the second magnon group and the external pumping field.


\begin{figure*}
  \includegraphics[]{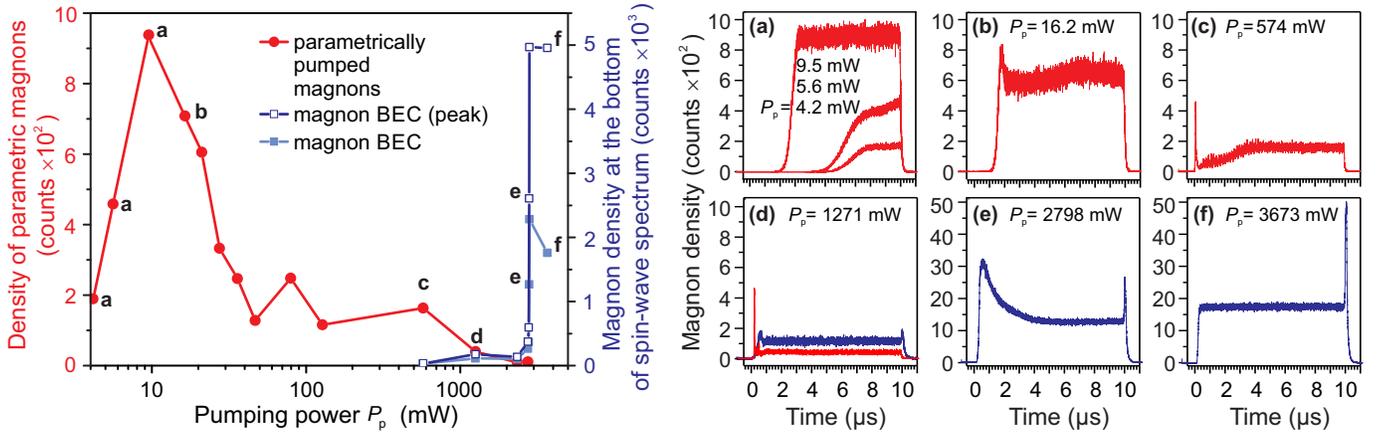}
  \caption{\label{Fig2}(color online). Dependence of the density of parametrically excited magnons at half of the pumping frequency $f_\mathrm{p}/2$ (red dots) and the density of the magnons at the bottom of the spin-wave spectrum (blue squares) on the pumping power $P_\mathrm{p}$. The magnon densities are measured as the intensities of the corresponding BLS signals. The selected temporal profiles of these signals are presented in panels (a)--(f). The pumping pulse of 10\,$\mu$s duration is applied at zero moment of time. The data marked by the red dots and by the filled blue squares were measured from the corresponding temporal profiles just before switching off the pump pulse. The open blue squares show the intensity of the peak of the magnon Bose-Einstein condensate, which is formed due to the evaporative supercooling \cite{Serga2014} of the magnon gas after the pumping is shut down [see e.g. panels (e)--(f)]. The letters near the data points relate them to the BLS temporal profiles in the corresponding panels. Three waveforms in panel \,(a) demonstrate the development of the conventional parametric instability with increasing pump power.}
\end{figure*}

The measurements were performed using a low-damping ferrimagnetic film of yttrium iron garnet (YIG, $\mathrm{Y_{3}Fe_{5}O_{12}}$) by means of a combined microwave and Brillouin light scattering (BLS) setup schematically shown in Fig.~\ref{Fig1}(a). The YIG film of $6.7\,\mu$m thickness with saturation magnetization $4 \pi M_\mathrm{s} = 1750$\,G was grown in the (111) crystallographic plane on a gallium gadolinium garnet substrate by liquid phase epitaxy. The in-plane magnetized YIG-film sample with dimensions ${2} \times 10\,\text{mm}^2$ was placed on top of a $50\,\mu$m wide microstrip resonator, which was used to induce the pumping Oersted field $\mathbf{h}(t)$ [see Fig.~\ref{Fig1}(a)].
The resonator was driven by microwave pulses at a carrier frequency of $ f_\text{p}=\omega_\text{p}/2\pi=13.74\,\text{GHz}$ with peak powers $P_\text{p}$ ranging from 1\,mW to 4\,W. The chosen pump pulse duration of $10\,\mu$s  was sufficiently long to observe the response of the spin-wave system even at the lowest applied pumping powers. A pulse repetition time of 100\,$\mu$s excluded microwave-heating effects.

In course of the parallel pumping process the photons of the microwave pumping field split into magnon pairs with opposite wavevectors at half of the pumping frequency as illustrated in Fig.~\ref{Fig1}(b). The strength of the bias field $H=1735$\,Oe has been chosen to allow for a pumping of the magnon pairs slightly above the ferromagnetic resonance frequency (FMR). In this case, firstly, the parallel pumping achieves its highest efficiency because the magnons are pumped to the transversal spectral branch, which has the lowest threshold of the parametric instability \cite{Serga2012}. Secondly, no kinetic instability process, which corresponds to a one-step scattering of the parametric magnons to the bottom of the spin-wave spectrum \cite{Melkov1994} and which may disturb the formation of the magnon BEC, is allowed due to energy and momentum conservation \cite{Melkov1991}.

The magnon system was analyzed by means of time-resolved BLS  spectroscopy \cite{Buttner2000} with a time resolution of 1~ns. A probing laser beam with a power of 5~mW and a wavelength of 532~nm was focused to a focal spot with a diameter of $50\,\mu$m in the middle of the microstrip resonator. The optics to focus the BLS laser onto the YIG sample was also used to collect the scattered light, which was analyzed by a multipass tandem Fabry-P\'{e}rot interferometer. The magnons at the frequency $f_\mathrm{p}/2$ and at the bottom of the spin-wave spectrum were detected simultaneously. The corresponding BLS signals were collected in 150~MHz wide frequency bands. The wavenumber sensitivity in the range $\pm 5\cdot10^4\,\mathrm{rad/cm}$ was insured by use of a wide-aperture optics.


Figure~\ref{Fig2} presents the dependencies of the magnon densities in the chosen spectral regions on the input pumping power. The red and blue waveforms show the time evolution of the parametric and the magnons at the bottom of the spectrum, respectively. One can see in Fig.~\ref{Fig2}(a) that the parametric magnons appear already at a small pumping power of 4.2\,mW. The increase of the pumping power to 6.7\,mW and to 9.6\,mW results in an earlier appearance and in a faster saturation of the parametric magnons at higher density levels as shown in Fig.~\ref{Fig2}(a). In accordance with the theory of the parametric spin-wave  instability \cite{Gurevich1996} this saturation is caused by a nonlinear dephasing of the excited magnons relative to the microwave pumping field $\mathbf{h}(t)$. 

Surprisingly, a further increase of the pumping power leads to a consecutive \textit{decrease} of the saturated magnon density. At a pumping power $P_\mathrm{p}\simeq 1270$\,mW, when the thermalized magnons start to be detected at the bottom of the spin-wave spectrum, the parametric magnons are hardly visible anymore [see waveforms in Fig.~\ref{Fig2}(d)]. For pumping powers above 2770\,mW the parametric magnons are practically non-detectable with the exception of a sharp initial peak similar to one in Fig.~\ref{Fig2}(d). Nevertheless, at $P_\mathrm{p} \geq 2770$\,mW the density of the low-energy magnons sharply rises up due to the formation of the Bose-Einstein condensate. 



Conventional four-magnon scattering processes, which were considered in the past to be responsible for the thermalization of a magnon gas and, thus, for the formation of the magnon BEC, cannot be assumed as a reason for the disappearance of the parametrically pumped magnon group. During the action of the external pumping this nonlinear mechanism can only limit but not reduce the number of the parametric magnons: The decrease in their density would lead to the consequent decrease of the scattering efficiency and, thus, to the density saturation.


The phenomenon can be understood if one takes into account the role of the external pumping field in the four-magnon scattering process. The time-dependent evolution of a spin wave with amplitude $c_q$, wavevector  $\mathbf{q}$, and complex frequency $\omega_q=\omega'_q+i\Gamma_q$ can be described by the following equation \cite{Gurevich1996}, where $\Gamma_q$ is the spin-wave relaxation frequency:
\begin{equation} \label{eq1}
\begin{split}
\frac{d c_q}{d t} = & \, i \omega_q c_q + iV_q h e^{i\omega_\mathrm{p} t} c^\ast_{-q} \\
& + i\sum_{q'} S_{qq'} c^{\ast}_{-q} c^{\vphantom{\ast}}_{q'} c^{\vphantom{\ast}}_{-q'} 
+ 2i\sum_{q'} T_{qq'} c^{\vphantom{\ast}}_{q} c^{\vphantom{\ast}}_{q'} c^{\ast}_{q'} 
\,.
\end{split}
\end{equation}
Here the first third-order term describes the interaction of magnon pairs $(c_{q}, c_{-q})$ and $(c_{q'}, c_{-q'})$. The second third-order term results in a frequency shift caused by the decreasing of $4 \pi M_\mathrm{s}$ due to the increase of the number of parametric
magnons.
The coupling between the magnon pair $(c_q, c_{-q})$ and the pumping field $h$ is represented by the parameter $V_q$:
\begin{equation} \label{eq2}
V_q=\frac{\gamma \omega_M}{4\omega_q} e^{i2\varphi_q} \sin^2\theta_q  \,,
\end{equation}
where $\omega_{M}=\gamma 4\pi M_\mathrm{s}$, $\gamma=2.8$\,MHz/Oe, $\theta_q=\angle(\mathbf{q}, \mathbf{H})$ and $\varphi_q=\angle(\mathbf{q}, \mathbf{q_{\perp}})$ are the polar and the azimuthal angles of the spin wave $c_q(\theta_q, \varphi_q)$, respectively. According to Eqs~(\ref{eq1}, \ref{eq2}) the magnons with $\theta_q=\theta_{1}=\pi/2$ possess the minimal threshold of the parametric generation $h=h_\mathrm{th}$, and thus are generated first, exactly as it was observed in our experiment.

Above the threshold ($h>h_\mathrm{th}$), the stationary amplitude of the initially generated waves $c^\infty_1 = c^\infty_q(\frac{\pi}{2}, 0) = c^\infty_{-q}(\frac{\pi}{2}, \pi)$ is determined by the action of the pumping field $h$ and by the self-action of these waves:
\begin{equation} \label{eq3}
|c^\infty_1|^2=\dfrac{N_1}{2}=\frac{\sqrt{(h V_1)^2-\Gamma_1^2}}{2S_{11}} \,,
\end{equation}
where $S_{11}$ is the coefficient $S_{qq'}$ for the waves of the initially excited parametric group $c_1$. 
Thus, the total pumping field in the sample consists of the external pumping field $h$ and an internal pumping field induced by the magnon group $c_1$. This pumping acts not only on the waves $c_1$ but on all other waves in the system. If its value exceeds the excitation threshold of any wave with $\theta \neq \pi/2$ this second group $c_2$ will be excited. 

After the excitation of the second magnon group the total pumping will consist of three components: The action of the external pumping field and the reactions of the first and the second groups. 
The effects of this complex pumping scenario are manifold. For example, the pumping can lead to the excitation of a third magnon group. It can also lead to the \textit{suppression} of the first group of parametric magnons. The total pumping that drives the first group equals
\begin{equation} \label{eq4}
\begin{split}
|P_1|^2 = & \, (h V_1)^2 +(S_{11}N_1+S_{21}N_{2})^2 \\
& -2(S_{11}N_1+S_{21}N_{2})\sqrt{(h V_1)^2-\Gamma_1^2} \,,
\end{split}
\end{equation}
where the product of $N_{2}=2|c^\infty_2|^2$ and $S_{21}$ describes the action of the second group on the first one. The minus sign before the last term in Eq.~\ref{eq4} denotes the decrease of the pump acting on the first group.
If $|P_1|^2 < |\Gamma_1|^2$ the first group will decay and only the second group will remain in course of time. This process can be understood as a \emph{stimulated scattering} of the initially injected magnons to a new spectral area. 


In order to verify the proposed model we performed an additional experiment on detection  of the second magnon group using the angle-resolving BLS setup shown in Fig.~\ref{Fig1}(a). The wavevector selection was realized by changing the angles $\Theta_{q\perp}$ and $\Theta_{q\parallel}$ between the sample and the incident light beam \cite{Sandweg2010}. It allowed us to increase the detectable wavenumber range up to $1.5\cdot10^5\,\text{rad/cm}$ for both  transversal and longitudinal magnons. The wavevector resolution was set to $2\cdot10^3\,\text{rad/cm}$ by a pinhole in the path to the interferometer. To increase the intensity of the magnon-dependent BLS signal the separation between pumping pulses was decreased to $5\,\mu$s, which was still sufficiently long to ensure the relaxation of the magnon system to the ground state after each pumping event. Accordingly, the duration of the pump pulses was decreased to $100\,\text{ns}$ and the measurements were performed at the maximal pump power. 

\begin{figure}
  \includegraphics[]{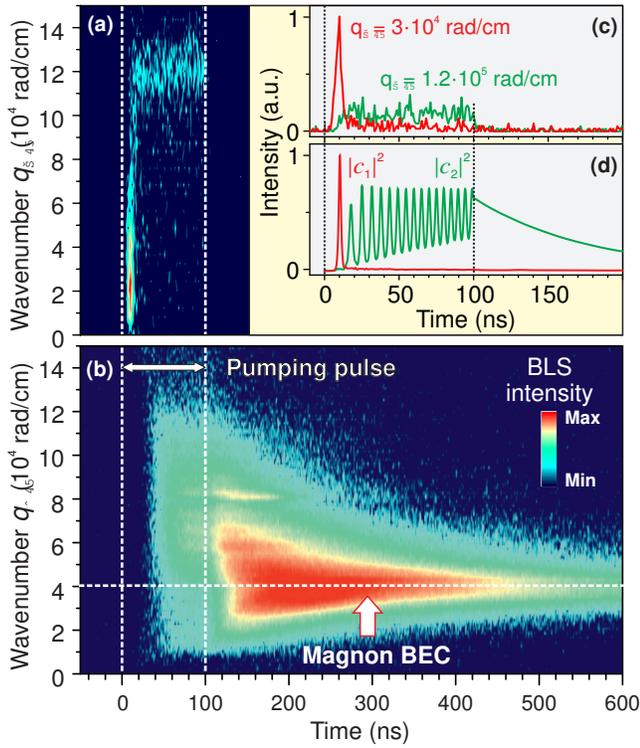}
  \caption{\label{Fig3}(color online). Color-coded logarithmic BLS intensity maps of (a) parametrically injected magnons ($\bf{q}\perp\bf{H}$) at $f_\text{p}/2$ and (b) the magnon gas ($\bf{q}\parallel\bf{H}$) at the bottom of the spin-wave spectrum. (c) Change in the density of the first (red curve) and the second (green curve) magnon groups extracted from the intensity map in panel (a) for the given wavenumbers $q_\perp$. (d) Time-dependent variations in the densities $|c_1|^2$ and $|c_2|^2$ of the first and the secondary  groups of parametric magnons numerically calculated using Eq.~(\ref{eq1}). The long decay tail of the second magnon group $c_2$ is due to neglecting the nonlinear decay caused by the conventional four-magnon scattering processes. The intensity oscillations are caused by a periodic dephasing between this magnon group and the total pumping field.}
\end{figure}

Figure~\ref{Fig3}(a) shows the color-coded BLS intensity map measured in time and wavevector space in the frequency range from $6.5\,\text{GHz}$ to $7.5\,\text{GHz}$ for magnons with $\bf{q}\perp\bf{H}$. The two dashed lines indicate the $100\,\text{ns}$ long microwave pumping pulse. The parametrically injected magnons with a frequency of $f_\text{p}/2$ appear first at wavenumbers in the range from $1.5\cdot10^4\,\text{rad/cm}$ to $3\cdot10^4\,\text{rad/cm}$ (see the red spot in the intensity map). Such a broad excitation range is due to the flatness of the perpendicular magnon branch near the FMR frequency (see Fig.~\ref{Fig1}(b), where the shadowed red segment covers the wavenumber range of the initially excited magnons). In the course of the next approximately $5\,\text{ns}$, some of the injected magnons redistribute towards smaller and higher wavenumbers. The short scattering time is a consequence of the high rate of magnon-scattering processes due to the large population of the primary group of magnons. Similar to the previous experiment [see Fig.~\ref{Fig2}(c) and (d)], the BLS signal from the parametrically injected magnons rapidly decreases and, approximately $20\,\text{ns}$ after the start of the pumping pulse, the first magnon group has completely vanished.  
At the same time, a weak BLS signal, which appears at wavenumbers between 
$1.10\cdot10^5\,\text{rad/cm}$ and $1.35\cdot10^5\,\text{rad/cm}$ and which is visible up to the end of the pump pulse, clearly provide evidence for the excitation of the second magnon group. It is noteworthy that the dynamics of the magnon groups $c_1$ and $c_2$ [see Fig.~\ref{Fig3}(d)], which was numerically calculated using Eq.~(\ref{eq1}) neglecting their interaction with all other modes of the magnon gas, qualitatively corresponds to the experimental time-dependent profiles of magnon densities shown in Fig.~\ref{Fig3}(c) as well in Figs~\ref{Fig2}(c)-(d).


Let us address now the formation of the magnon BEC and specifically a question regarding the influence of the second magnon group on the transition of thermalized magnons to the low-energy states. Figure~\ref{Fig3}(b) represents a wavenumber-resolved BLS intensity map of magnons with $\bf{q}\parallel\bf{H}$ within the frequency range from $4.7\,\text{GHz}$ to $5.7\,\text{GHz}$, which includes the bottom of the magnon spectrum. One can see that the first thermalized magnons appear already 30\,ns after the rising flank of the pumping pulse. However, during the pump pulse the formation of the magnon BEC is prevented by the extremely high effective temperature of the parametrically driven magnon gas \cite{Serga2014}. 
After the end of the pumping, the magnon temperature rapidly decreases due to the evaporative supercooling process \cite{Serga2014} and magnons form the Bose-Einstein condensate in the global energy minimum at $4\cdot10^4\,\text{rad/cm}$ [see Fig.~\ref{Fig1}(b)]. 
By comparing Fig.~\ref{Fig3}(a) and Fig.~\ref{Fig3}(b) one can see that the appearance of thermalized magnons at the bottom of the spin-wave spectrum correlates with the excitation of the second group of parametric magnons with a perpendicular wavevector component $q_\perp$ beyond $1.1\cdot10^5\,\text{rad/cm}$. 

This correlation can be understood from the momentum conservation law. It is obvious that no magnon-magnon scattering process within the initially excited transversal magnon group is able to turn the magnon wavevector from $\mathbf{q} \perp \mathbf{H}$  to $\mathbf{q} \parallel \mathbf{H}$ orientation. At the same time, such wavevector conversion can easily occur if the second spin-wave group possesses a sufficiently large wavevector component along $\mathbf{H}$. 

It is hard to predict the polar angle of the second group in our case as no general solution of this problem exists for a magnetic film of finite thickness.
However, it is expected that this angle significantly differs from $\pi/2$.
For example,  
in an unbounded medium the secondary wave group could be excited under the angle $\theta_{2} \simeq 50^\circ$ \cite{Zautkin1972}. 
The situation is even more complicated for a spatially localized pumping realized as in our case by the narrow microstrip resonator: Because excited waves can leak from the pumping area their losses strongly depend on their group velocities $\mathbf{v}_\mathrm{gr}$. The first magnon group with $\theta_{1} = \pi/2$ propagates along the microstrip resonator and, thus, does not suffer any radiation losses. Practically all other magnons with $\theta_q \neq \pi /2$ can easily leak out of the $50\,\mu$m wide pumping area and their excitation thresholds rise up significantly \cite{Neumann2009}. However, there are magnons with $\theta_q \neq \pi /2$ which group velocities are still directed along the microstrip resonator. Four of such spectral positions ($n=12$, $\theta_q=35^\circ$, $q=2.16\cdot10^4\,\text{rad/cm}$, $q_\perp=\pm 1.2\cdot10^5\,\text{rad/cm}$) forming the second magnon group are marked by the green dots on the isofrequency plane $f_\mathrm{p}/2$  in Fig.~\ref{Fig1}(b). In general, a few thickness modes $n$ can be excited simultaneously as evidenced by the broad BLS signal produced by the second group in Fig.~\ref{Fig2}(a). As a result, the stimulated scattering brakes the orthogonality between the first magnon group ($\theta_q=90^\circ$) and magnons from the bottom of the spin-wave spectrum ($\theta_q=0$), and thus create the prerequisites for Bose-Einstein condensation. Our evaluation shows that the second magnon group can further scatter by the conventional four-magnon scattering between the neighboring modes while conserving energy and momentum. Such intermodal scattering processes are allowed down to the minimum of the first $n=0$ mode where the BEC is formed at $q=4\cdot10^4\,\text{rad/cm}$ and $\theta_q=0$ [see Fig.~\ref{Fig1}(b)]. 

In conclusion, our experimental findings suggest distinctly different phases of the thermalization process of a parametrically driven magnon gas:  First, energy degenerated, parametrically   magnons are generated, which, secondly, scatter in a multistage magnon transfer process to the bottom of the magnon spectrum. Thus, the parametrically stimulated magnon thermalization is assumed to be the essential precondition for the formation of the magnon Bose-Einstein condensate.

We thank T.~Br\"acher for fruitful discussions. Financial support from the Deutsche Forschungsgemeinschaft within the SFB/TR 49 and from the Ukrainian Fund for Fundamental Research is gratefully acknowledged.


\begin{thebibliography}{1}

\bibitem{Schloemann1960} E.~Schl\"omann, J.J.~Green, and U.~Milano,
J. Appl. Phys. {\bf 31}, 386S (1960).

\bibitem{Rezende1990} S.M.~Rezende and F.M.~de~Aguiar,
IEEE Proc. {\bf 78}, 6 (1990).

\bibitem{Demokritov2003} S.O.~Demokritov, A.A.~Serga, V.E.~Demidov, B.~Hillebrands, M.P.~Kostylev, and B.A.~Kalinikos, 
Nature {\bf 426}, 159 (2003).

\bibitem{Braecher2011} T.~Br\"acher, P.~Pirro, B.~Obry, B.~Leven, A.A.~Serga, and B.~Hillebrands,
Appl. Phys. Lett. {\bf 99}, 162501 (2011).

\bibitem{Kalinikos1994} B.A.~Kalinikos and M.P.~Kostylev,
IEEE Transactions on Magnetics {\bf 33}, 3445 (1994).

\bibitem{Bagada1997} A.V.~Bagada, G.A.~Melkov, A.A.~Serga, and A.N.~Slavin, 
Phys. Rev. Lett. {\bf 79}, 2137 (1997).


\bibitem{Braecher2014}	T.~Br\"acher, F.~Heussner, P.~Pirro, T.~Fischer, M.~Geilen, B.~Heinz, B.~L\"agel, A.A.~Serga, and B.~Hillebrands, 
Appl. Phys. Lett. {\bf 105}, 232409 (2014).

\bibitem{Melkov2001} G.A.~Melkov, Yu.V.~Kobljanskyj, A.A.~Serga, A.N.~Slavin, and V.S.~Tiberkevich, 
Phys. Rev. Lett. {\bf 86}, 4918 (2001).

\bibitem{Chumak2014} A.V.~Chumak, V.I.~Vasyuchka, A.A.~Serga, M.P.~Kostylev, V.S.~Tiberkevich, and B.~Hillebrands,
Phys. Rev. Lett. {\bf 108}, 257207 (2012).

\bibitem{Demokritov2006} S.O.~Demokritov, V.E.~Demidov, O.~Dzyapko, G.A.~Melkov, A.A.~Serga, B.~Hillebrands, and A.N.~Slavin,
Nature {\bf 443}, 430 (2006).

\bibitem{Pokrovsky2013} F.~Li, W.M.~Saslow, and V.L.~Pokrovsky, 
Sci. Rep. {\bf 3}, 1372 (2013).

\bibitem{Demidov2007} V.E.~Demidov, O.~Dzyapko, S.O.~Demokritov, G.A.~Melkov, and A.N.~Slavin,
Phys. Rev. Lett. {\bf 99}, 037205 (2007).

\bibitem{Demidov2008-1} V.E.~Demidov, O.~Dzyapko, M.~Buchmeier, T.~Stockhoff, G.~Schmitz, G.A.~Melkov, and S.O.~Demokritov, 
Phys. Rev. Lett. {\bf 101}, 257201 (2008).

\bibitem{Demidov2008-2} V.~Demidov, O.~Dzyapko, S.~Demokritov, G.~Melkov, and A.~Slavin,
Phys. Rev. Lett. {\bf 100}, 047205 (2008).

\bibitem{Chumak2009} A.~Chumak, G.~Melkov, V.~Demidov, O.~Dzyapko, V.~Safonov, and S.~Demokritov,
Phys. Rev. Lett. {\bf 102}, 187205 (2009).

\bibitem{Serga2014} A.A.~Serga, V.S.~Tiberkevich, C.W.~Sandweg, V.I.~Vasyuchka, D.A.~Bozhko, A.V.~Chumak, T.~Neumann, B.~Obry, G.A.~Melkov, A.N.~Slavin, and B.~Hillebrands,
Nat. Commun. {\bf 5}, 4452 (2014).

\bibitem{Hick2012}
J.~Hick, T.~Kloss, and P.~Kopietz, 
Phys. Rev. B {\bf 86}, 184417 (2012).

\bibitem{Kalinikos1986} B.A.~Kalinikos and A.N.~Slavin,
J. Phys. C: Solid State Phys. {\bf 19}, 7013 (1986).

\bibitem{Serga2012} A.A.~Serga, C.W.~Sandweg, V.I.~Vasyuchka, M.B.~Jungfleisch, B.~Hillebrands, A.~Kreisel, P.~Kopietz, and M.P.~Kostylev,
Phys. Rev. B {\bf 86}, 134403 (2012).

\bibitem{Melkov1994} G.A.~Melkov, V.L.~Safonov, A.Y.~Taranenko, and S.V.~Sholom, 
J. Magn. Magn. Mater. {\bf 132}, 180 (1994).

\bibitem{Melkov1991} G.A.~Melkov and S.V~Sholom, 
Sov. Phys. JETP {\bf 72}, 341 (1991).

\bibitem{Buttner2000} O.~B\"uttner, M.~Bauer, S.O.~Demokritov, B.~Hillebrands, Yu.S.~Kivshar, V.~Grimalsky, Yu.~Rapoport, and A.N.~Slavin,
Phys. Rev. B {\bf 61}, 11576 (2000).

\bibitem{Gurevich1996} A.G.~Gurevich and G.A.~Melkov, 
\textit{Magnetization Oscillations and Waves} (CRC Press, New York, 1996).

\bibitem{Sandweg2010} C.W.~Sandweg, M.B.~Jungfleisch, V.I.~Vasyucka, A.A.~Serga, P.~Clausen, H.~Schultheiss, B.~Hillebrands, A.~Kreisel, and P.~Kopietz,
Rev. Sci. Instrum. {\bf 81}, 073902 (2010).



\bibitem{Zautkin1972} V.V. Zautkin, V.E. Zakharov, V.S. L’vov, S.L. Musher, and S.S. Starobinets, 
Sov. Phys. JETP {\bf 35}, 926 (1972).


\bibitem{Neumann2009} T.~Neumann, A.A.~Serga, V.I.~Vasyuchka, and B.~Hillebrands, 
Appl. Phys. Lett. {\bf 94}, 192502 (2009). 


\end{thebibliography}
\end{document}